\documentclass[prl,aps,twocolumn,notitlepage,10pt]{revtex4-1}
\usepackage[hidelinks]{hyperref} 
\usepackage{lmodern}
\usepackage{amsmath}
\usepackage{amssymb}
\usepackage{epsfig}

\def\d{\mathrm{d}}
\newcommand{\be}{\begin{equation}}
\newcommand{\ee}{\end{equation}}
\newcommand{\beA}{\begin{equation}\begin{aligned}}
\newcommand{\eeA}{\end{aligned}\end{equation}}

\begin{document}
\title{Reservoir-induced stabilisation of a periodically driven many-body system}
\date{\today}
\author{Thomas Veness}
\affiliation{School of Physics and Astronomy, University of Nottingham,
  Nottingham, NG7 2RD, United Kingdom}
\author{Kay Brandner}
\affiliation{School of Physics and Astronomy, University of Nottingham,
  Nottingham, NG7 2RD, United Kingdom}
\begin{abstract}
Exploiting the rich phenomenology of periodically-driven many-body systems is
notoriously hindered by persistent heating in both the classical and quantum
realm.
Here, we investigate to what extent coupling to a large thermal reservoir makes
stabilisation of a non-trivial steady state possible.
To this end, we model both the system and the reservoir as classical spin
chains where driving is applied through a rotating magnetic field, and simulate
the Hamiltonian dynamics of this setup. 
We find that the intuitive limits of infinite and vanishing frequency, where
the system dynamics is governed by the average and the instantaneous
Hamiltonian, respectively, can be smoothly extended into entire regimes
separated only by a small crossover region. 
At high frequencies, the driven system stroboscopically attains a Floquet-type
Gibbs state at the reservoir temperature. 
At low frequencies, a synchronised Gibbs state emerges, whose temperature may
depart significantly from that of the reservoir.
Although our analysis in some parts relies on the specific properties our
setup, we argue that much of its phenomenology should be generic for a large
class of systems.
\end{abstract}

\maketitle
\addcontentsline{toc}{section}{Introduction}
\paragraph{Introduction.--}
Our understanding of periodically-driven quantum systems is largely
facilitated by \emph{Floquet's theorem}
\cite{Kitagawa-Oka-Brataas-Fu-Demler,Sambe,Torres-Kunold}, which implies that
the explicit time-dependence of Schr\"{o}dinger's equation can be removed by
means of a unitary basis transformation with the same periodicity as the drive
\cite{Bukov-DAlessio-Polkovnikov-review,Eckardt-rmp}.
Hence, the stroboscopic time evolution of a such systems is generated by a
time-independent Hermitian operator, the Floquet Hamiltonian, whose properties
can be tailored through the applied driving protocol.
This result opened the field of Floquet engineering \cite{Weitenberg-Simonet}
and led to the discovery of novel phenomena with no static counterpart such as
anomalous Floquet topological insulators
\cite{Rubio-Abadal,Cayssol-Dora-Simon-Moessner} or so-called time crystals
\cite{Else-Bauer-Nayak,vonKeyserlingk-Khemani-Sondhi,Sacha-Zakrzewski}.

The Floquet Hamiltonian of a many-body system is generally a complicated
object, which in most cases can be determined only via approximate methods.
Its eigenstates, which govern the long-time behaviour of the system, are
generically superpositions of its instantaneous energy eigenstates with
quasi-homogeneously distributed coefficients 
\cite{Moessner-Sondhi,Zhang-Khemani-Huse}.
Combining this observation with the notion that energy eigenstates typically
display thermal behaviour, i.e. the \emph{eigenstate thermalisation hypothesis}
\cite{Deutsch,Srednicki,Rigol-Dunjko-Olshanii,DAlessio-Kafri-Polkovnikov-Rigol},
leads to an ensemble with no conservation laws
\cite{DAlessio-Rigol,Kim-Ikeda-Huse}.
That is, a generic many-body system continuously absorbs energy until the
statistics of all observables are described by a structureless
\emph{infinite-temperature ensemble}
\cite{Ponte-Chandran-Papic-Abanin,Russomanno-Silva-Santoro,Kuwahara-Mori-Saito,Abanin-DeRoeck-Huveneers}.
In practice, this behaviour confines Floquet engineering to the high-frequency
regime, where heating is exponentially suppressed, or to special systems where
thermalisation is hindered by other means such as many-body localisation
\cite{Fleckenstein-Bukov,Ishii-Kuwahara-Mori-Hatano,BAA,Nandkishore-Huse,vonKeyserlingk-Sondhi}.

Floquet theory does not apply to systems obeying non-linear dynamics.
A variety of techniques used in analysing periodically driven quantum systems
such as high-frequency expansions may, however, be naturally adopted in a
classical Hamiltonian framework \cite{Bukov-DAlessio-Polkovnikov-review} and
much phenomenology may persist.
\begin{figure}[h!]
  \includegraphics[width=0.8\linewidth]{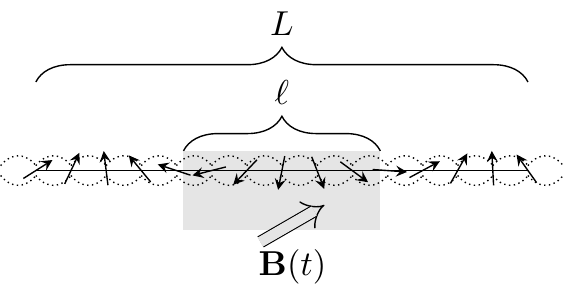}
  \includegraphics[width=\linewidth]{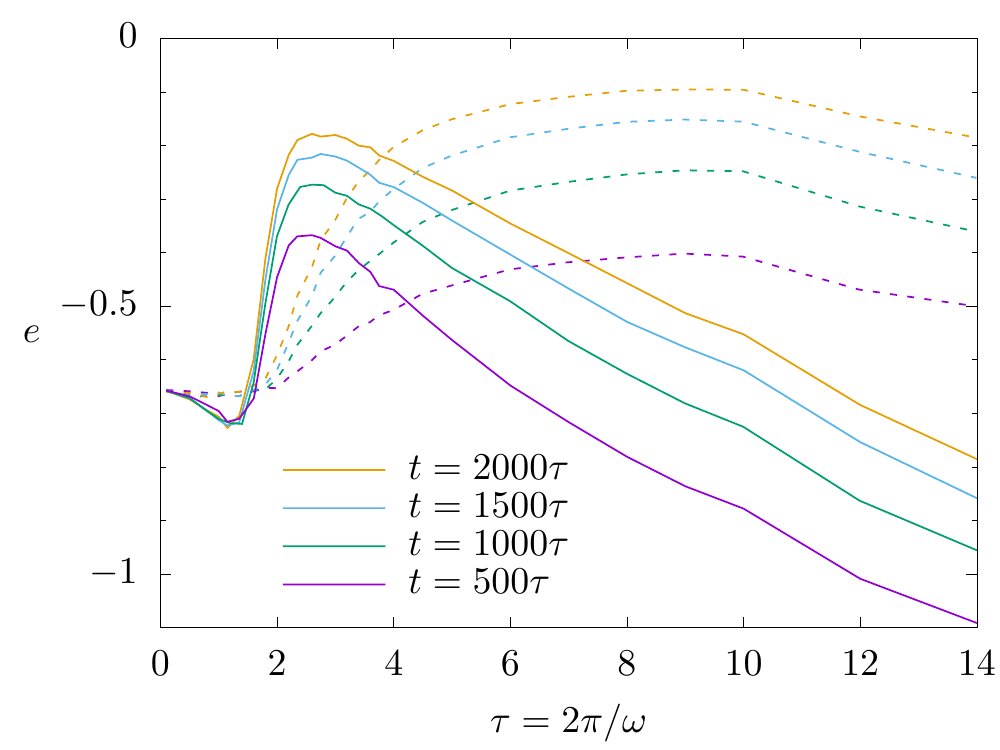}
  \caption{
    Above: Sketch of the system.
    A classical spin chain with nearest-neighbour interactions, wark disorder,
    and total number of sites $L$, $\ell$ of which are exposed to a time-periodic
    magnetic field ${\bf B}(t)$.
    Below: Mean energy density of the driven system as a function of the period
    $\tau$ over the elapsed time $t$.
    While nearly no energy is absorbed for $\tau \lesssim 2$, a
    reservoir-stabilised steady state with strongly suppressed absorption emerges
    for $\tau \gtrsim 3$, before heating eventually prevails due to finite-size
    effects (solid, $L=2000$). By contrast, the isolated system rapidly heats up
    for $\tau \gtrsim 2$ (dashed, $L= \ell$).
    For all plots we have set $\ell=20$, $\delta J = 2\times 10^{-2}$, $e_{\rm
    initial}=-0.66$, see main text for symbols.
  } \label{fig:reservoir-comparison}
\end{figure}
Nonetheless, systematic investigations of classical systems have begun only
recently, despite greater numerical accessibility of larger systems sizes and
longer evolution times than in the quantum case
\cite{Nunnenkamp,Howell,Mori-Floquet-prethermal,Citro-Dalla-Torre,Bukov-DAlessio-Polkovnikov-review,McRoberts-Bilitewski-Haque-Moessner}.
Here, we exploit this advantage to explore \emph{reservoir-induced
  thermalisation} as a new approach to avoid overheating in periodically driven
many-body systems.

For concreteness, we consider a classical spin chain with a magnetic drive
applied to a finite region, and the remainder of the system acting as a
reservoir, see Fig.~\ref{fig:reservoir-comparison}.
At high driving frequencies, the system is well-described by the time-averaged
Hamiltonian, up to perturbative corrections in the inverse driving frequency; 
this behaviour is qualitatively insensitive to the presence of the reservoir.
In the quasi-static limit, the system relaxes to the instantaneous Hamiltonian
before it changes appreciably.
At any finite frequency, however, the existence and nature of a stroboscopic
steady state is a priori unclear.
Our main finding is that the reservoir stabilises a non-trivial steady state
over a wide range of frequencies and on an intermediate but practically long
timescale, where the isolated system would overheat, see
Fig.~\ref{fig:reservoir-comparison}.
This steady state, which is smoothly connected to the quasi-static limit,
emerges through synchronisation between the driven system and the reservoir,
suppressing net energy absorption \cite{long-paper}, and is described by a
Gibbs ensemble whose temperature can deviate significantly from the initial
temperature of the reservoir.
This behaviour is in clear contrast with the intuitive expectation of the
reservoir acting as a mere energy sink, whose state is invariant up to
finite-size corrections.
In the following, we explicitly construct both high- and low-frequency
ensembles, which describe all local observables of the driven system, leaving
only a small crossover region unexplained by a steady state.
\addcontentsline{toc}{section}{Setup}
\paragraph{Setup.--}\label{sec:system}
We consider the setup of Fig.~\ref{fig:reservoir-comparison}.
We will refer to the $\ell$ driven sites as the \emph{system proper}, and the
remaining $L-\ell$ sites of the spin-chain as the \emph{reservoir}, where we
are particularly interested in the regime $\ell \ll L$.
The dynamical variables are classical spin vectors normalised such that $|{\bf
  S}_j|^2 = 1$.
The Hamiltonian of the entire system is
\be
H(t) = - \sum_{j=1}^L {\bf S}_j^\top J_j {\bf S}_{j+1} + \sum_{j=1}^\ell {\bf B}(t) \cdot {\bf S}_j.
\label{eq:H}
\ee
We assume periodic boundary conditions i.e.  ${\bf S}_{L+1} = {\bf S}_1$.
The $J_j$ are diagonal matrices whose entries $J_j^\alpha$ are independently
and identically distributed and drawn from a normal distribution with mean
$J=1$, which we use as our energy scale throughout, and variance $\delta J$.
The driving ${\bf B}(t) = J(\cos \omega t, \sin \omega t,
  0)^\top$ is a rotating planar magnetic field with period $\tau = 2\pi /
  \omega$.
We choose $\delta J$ small but non-zero to ensure that there are no conserved
quantities, which would constrain the dynamics.

The time evolution of the system is determined by Hamilton's equations of
motion, $\frac{\d f}{\d t} = \frac{\partial f}{\partial t} + \{ f, H\}$, where
the Poisson bracket for spin degrees of freedom is determined by the Lie
algebra of $SO(3)$
\be
\{ S_j^\alpha, S_k^\beta \} = \delta_{jk} \varepsilon^{\alpha\beta\gamma} S_j^\gamma.
\ee
Accordingly, the microscopic equations of motion are
\beA
\frac{\d {\bf S}_j }{\d t}
& = -{\boldsymbol\Omega}_j \times {\bf S}_j , \\
{\boldsymbol\Omega}_j &=
J_{j-1} {\bf S}_{j-1}
+
J_j {\bf S}_{j+1}
-
\begin{cases} {\bf B}(t) & 1\leq j \leq \ell \\
  0          & \ell < j \leq L\end{cases}
.
\eeA
We choose our initial state as a global equilibrium state with zero magnetic
field.
Initial conditions are sampled using standard Metropolis-Hastings
Monte Carlo (MC) methods from the Gibbs distribution
\be
P_0(\{ {\bf S}_j \} ) =
e^{-\beta H_0}/Z_0
\label{eq:P0}
\ee
with $H_0=- \sum_{j=1}^L {\bf S}_j^\top J_j {\bf S}_{j+1}$
\cite{Newman-Barkema}.
Throughout this article, $P$ will denote a probability density and $Z$ a
normalisation constant.
The dynamics are numerically realised via symplectic
integration which manifestly conserves
spin-normalisation \cite{Krech-Bunker-Landau}.
Unless otherwise stated, presented results are averages over a large number of
initial states and disorder realisations.
We confirm in Ref.~[\onlinecite{long-paper}] that in the absence of driving and
with small $\delta J =10^{-3}$, time and ensemble averages are indeed equivalent
i.e. the free system is ergodic.

We first explore the existence of a steady-state through observables of the
system proper, specifically focusing on magnetisation and energy density, given
by
\beA
{\bf m} &= \frac{1}{\ell} \sum_{j=1}^\ell {\bf S}_j ,\\
e &= - \frac{1}{\ell-1} \sum_{j=1}^{\ell-1} {\bf S}_j^\top J_j {\bf S}_{j+1} +  {\bf m} \cdot {\bf B}(t) . \label{eq:emdef}
\eeA
The energy of the system proper is a salient observable through which to
explore reservoir-induced stabilisation.
In Fig.~\ref{fig:reservoir-comparison}, we plot the stroboscopic energy density
$e(t=n\tau)$ after a large number of cycles $n$, where we set $\delta J = 0.02$
such that heating is observable on the timescales considered.
To gain a clearer theoretical picture of the different regimes, we henceforth
take $\delta J = 10^{-3}$ in order to increase the separation of timescales
between heating and relaxation to the intermediate steady state.
\begin{figure*}[ht!]
  \includegraphics[width=\linewidth]{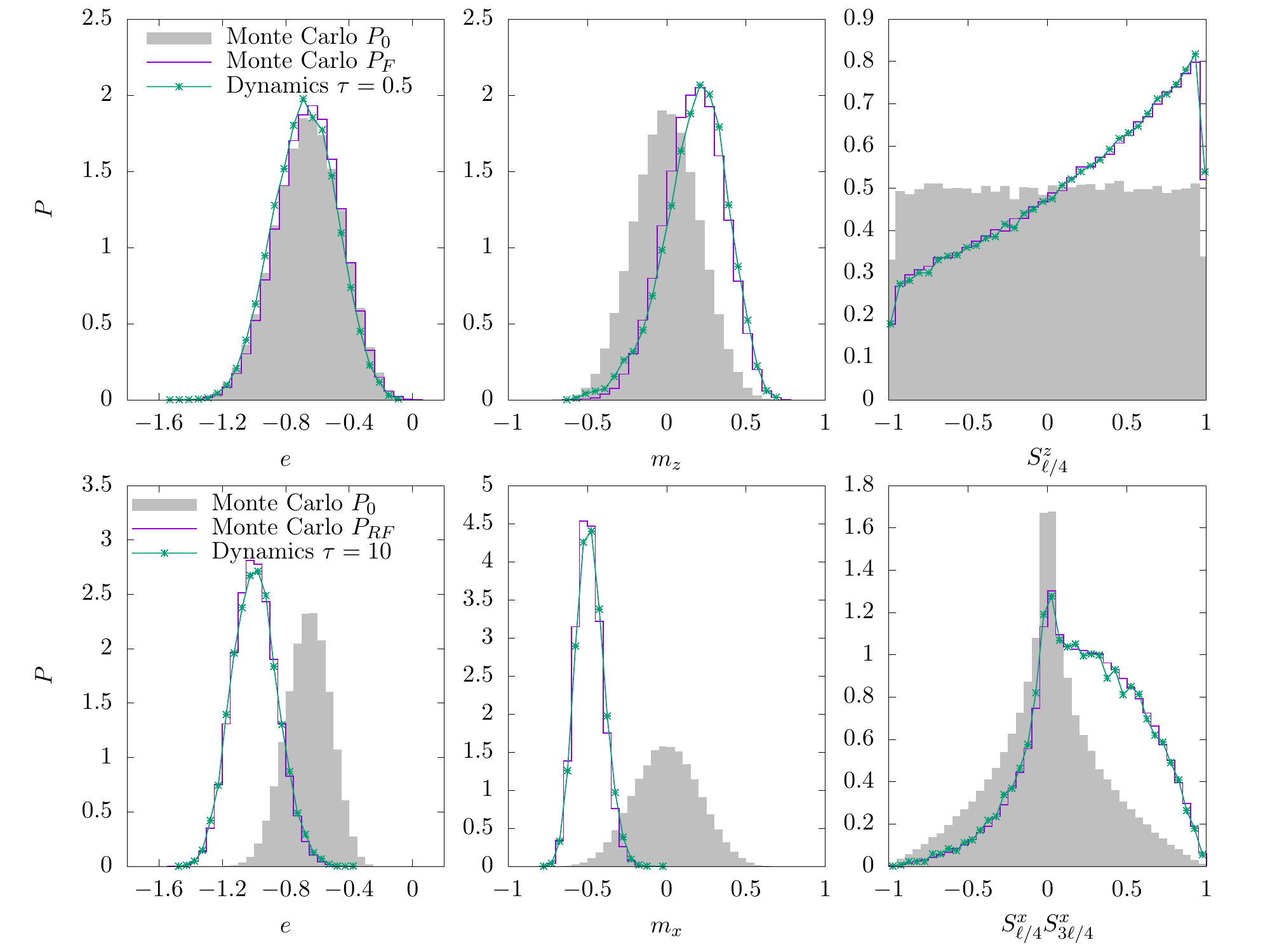}
  \caption{
  Probability density histograms of stroboscopic
  observables for dynamical evolution and statistical
  descriptions, $\ell=20$,
  $L=2000$, $\delta J = 10^{-3}$, $\beta$ chosen such
  that $e_{\rm initial}=-0.66$.
  Here $e$, $m_x$, $m_z$, and $m_x$ are determined by
  Eq.~(\ref{eq:emdef}), $S^z_{\ell/4}$ is the $z$-component of the
  spin at site $j=\ell/4$, and $S^x_{\ell/4} S^x_{3\ell/4}$ is
  the product of the $x$-components of the spins at sites $j=\ell/4$ and
  $j=3\ell/4$, highlighting that spatial correlations are also captured by the
  stroboscopic low-frequency ensemble of Eq.~(\ref{eq:low-freq-gibbs}).
  For comparison, the initial ensemble $P_0$ of Eq.~(\ref{eq:P0}) is
  shown in grey.
  Top: fast-driving regime, $\tau=0.5$, purple curves show MC sampling of from
  Eq.~(\ref{eq:floquet-gibbs}) with $\beta_{F} = \beta$, dynamics sampled at
  $t=2000\tau$.
  Bottom: slow-driving regime, $\tau=10$, where purple curves show MC
  sampling of from Eq.~(\ref{eq:low-freq-gibbs}) with $\beta_{RF}$ determined
  from Eq.~(\ref{eq:betarf-stat}), and dynamics are sampled at $t=1000\tau$.
  }
  \label{fig:histogram-array}
\end{figure*}

\addcontentsline{toc}{section}{High-frequency regime}
\paragraph{High-frequency regime.--}\label{sec:high-frequency}
When the system is driven at a frequency $\omega \gg J$
we expect its physics to be well-approximated by the time-averaged Hamiltonian,
with corrections scaling as powers of $\omega^{-1}$, given by the classical
Floquet-Magnus expansion \cite{Oteo-Ros, Blanes-Casas-Oteo-Ros,
large-omega-comment, Bukov-DAlessio-Polkovnikov-review}.
Note that we formally set $\hbar=1$ such that the spins are dimensionless and
energies and inverse timescales have the same dimension.
Such expansions are generally asymptotic in $\omega^{-1}$ and only
in special cases is resummation possible \cite{feldman}.
More precisely, rigorous recent results  on the classical limit of quantum
spin-chains demonstrate that the Floquet-Magnus Hamiltonian approximates a
conserved quantity as $\omega\to\infty$, although the stroboscopic evolution of
observables generated by $H_{F}$ is not guaranteed to converge to those of the
true time-dependent evolution \cite{Mori-Kuwahara-Saito-prl-2016}.
In our case, the leading order gives the Floquet Hamiltonian
\be
H_{F}^{(1)} = - \sum_{j=1}^L {\bf S}_j^\top  J^{HF}_j {\bf S}_{j+1} 
- \frac{J^2}{2\omega}\sum_{j=1}^\ell {\bf \hat{z}} \cdot {\bf S}_j 
+ \mathcal{O}(\omega^{-2}) \label{eq:hf1}
\ee
with the modified couplings $J^{HF}_j = Q_j^\top J_j Q_{j+1}$, where $Q_j=
R_y(J/\omega)$ represents rotation around the $y$-axis by $J/\omega$ for
$j=1,\ldots,\ell$ and the identity otherwise
\cite{Bukov-DAlessio-Polkovnikov-review}.

In the absence of other conserved quantities, statistical-mechanics leads us to
posit a stroboscopic Floquet-Gibbs ensemble  of the form
\be
P_{F}(\{{\bf S}_j\}) = e^{-\beta_{F} H_{F}}/Z_{F},
\label{eq:floquet-gibbs}
\ee
where $H_{F}$ is an appropriately truncated Floquet-Magnus expansion.
For our purposes, we find it sufficient to set $H_{F} = H_{F}^{(1)}$
\cite{Floquet-gibbs-1,Floquet-gibbs-2}.
As $\omega \gg J$, energy absorption is significantly suppressed, and we expect
that the effective inverse temperature fulfills $\beta_{F} \approx \beta$.
Fig.~\ref{fig:histogram-array} shows that such an ensemble reproduces the
statistics of system proper observables well for $\tau=0.5$, where the leading
correction in $\omega^{-1}$ is necessary to capture the asymmetry in
observables involving $S^z_j$.
We confirm in Ref.~[\onlinecite{long-paper}] that higher-order corrections can
indeed be safely neglected.

\addcontentsline{toc}{section}{Low-frequency regime}
\paragraph{Low-frequency regime.--} \label{sec:low-frequency}
We observe that our model admits an almost-static description in the
rotating frame defined by introducing new variables
${\bf S}_j = R(t)\widetilde{\bf S}_j$,
where
\be
R(t) = 
\begin{pmatrix} 
\cos (\omega t) & -\sin (\omega t) & 0 \\ 
\sin (\omega t) & \cos (\omega t) & 0 \\ 
0 & 0 & 1
\end{pmatrix}.
\label{eq:rt}
\ee
The microscopic equations of motion become
\beA
\frac{\d \widetilde{\bf S}_j}{\d t}
&= - \widetilde{\boldsymbol\Omega}_j \times \widetilde{\bf S}_j + \mathcal{O}(\delta J)
= \{ \widetilde{\bf S}_j, H_{RF} \} + \mathcal{O}(\delta J),\\
\widetilde{\boldsymbol\Omega}_j &= J \widetilde{\bf S}_{j-1}
+ J \widetilde{\bf S}_{j+1}
+ \begin{cases}  \omega \hat{\bf z} - J \hat{\bf x}  & 1\leq j \leq \ell \\ \omega \hat{\bf z} & \ell < j \leq L \end{cases}.
\eeA
Here, we have neglected time-dependent corrections of order $\delta J$ and
introduce the rotating-frame Hamiltonian
\be
H_{RF} = -\sum_{j=1}^L \left[\widetilde{\bf S}_j \cdot \widetilde{\bf S}_{j+1}
  + \omega\hat{\bf z}\cdot \widetilde{\bf S}_j \right]
+ \sum_{j=1}^\ell \hat{\bf x} \cdot \widetilde{\bf S}_j
.
\ee
Since the transformation (\ref{eq:rt}) reduces to the identity at every integer
multiple of the period, the autonomous dynamics generated by $H_{RF}$ are
stroboscopically equivalent to the dynamics of the drive system in the limit
$\delta J \ll J$.

In this global Floquet picture, statistical mechanics implies that the total
system should, after some transient phase, be stroboscopically described by a
canonical ensemble
\be
P_{RF}(\{{\bf S}_j\}) = e^{-\beta_{RF} H_{RF}}/Z_{RF},
\label{eq:low-freq-gibbs}
\ee
where $\beta_{RF}$ is an effective inverse temperature.
Two remarks are in order here.
First, although we neglect order $\delta J$ corrections to the rotating-frame
Hamiltonian, which can be calculated systematically by means of a
Floquet-Magnus expansion \cite{long-paper}, we still assume that the disorder
makes the system ergodic.
Second, as the total system is closed, one would in principle have to use a
micro-canonical ensemble in Eq.~(\ref{eq:low-freq-gibbs}). However, here we
prefer the technically simpler canonical ensemble, which should be equivalent
for any local observables if $L\gg 1$.

On a timescale slower than the inverse heating rate $H_{RF}$ is approximately
conserved, and so the Lagrange multiplier $\beta_{RF}$ may be fixed by
requiring that the expectation value of $H_{RF}$ in the final ensemble of
Eq.~(\ref{eq:low-freq-gibbs}) is the same as in the initial ensemble of
Eq.~(\ref{eq:P0}) i.e.
\beA
\int \d {\bf S}_1 \dots \d {\bf S}_L
H_{RF}(\{ {\bf S}_j \} )
e^{-\beta H_0(\{ {\bf S}_j\})}/Z_0
=\\
\ \int \d {\bf S}_1 \dots \d {\bf S}_L
H_{RF}(\{ {\bf S}_j \} )
e^{-\beta_{RF} H_{RF}(\{{\bf S}_j \}) }/Z_{RF}
.
\label{eq:betarf-stat}
\eeA
In the limit $L\gg\ell$, these quantities are dominated by the reservoir terms
and become independent of $\ell$.
Thus Eq.~(\ref{eq:betarf-stat}) implicitly defines an effective reservoir
temperature $\beta_{RF} = \beta_{RF}(\beta, \omega)$, which may be found
numerically using MC methods.
We recover instantaneous equilibration in the quasi-static limit i.e.
$\lim_{\omega\to0} \beta_{RF}(\beta,\omega) = \beta$.

\begin{figure}[hb!]
  \includegraphics[width=\linewidth]{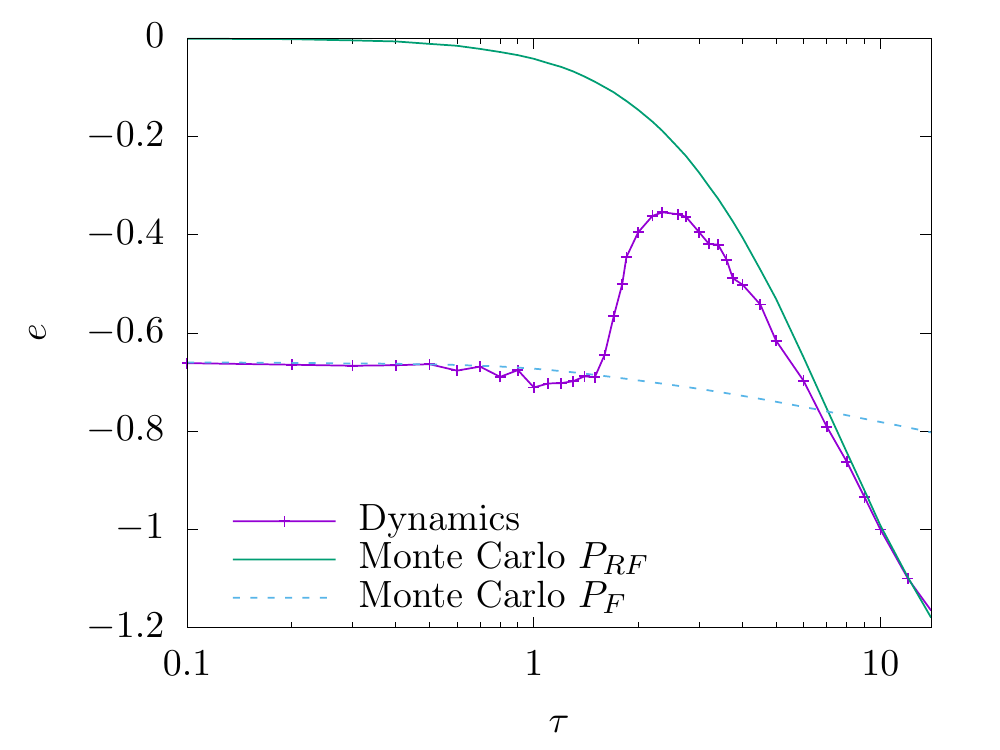}
  \caption{Energy of the system proper from dynamical simulations and
    ensemble descriptions of
    Eq.~(\ref{eq:floquet-gibbs}) with $\beta_{F} = \beta$ (high
    frequency), and
    Eq.~(\ref{eq:low-freq-gibbs}) (low frequency), with $\beta_{RF}$
    determined from Eq.~(\ref{eq:betarf-stat}).
    We have set $e_{\rm initial}=-0.66$, $\ell=20$, $L=2000$, $t=1000\tau$, $\delta J = 10^{-3}$.
  }
  \label{fig:beta-comp}
\end{figure}

We show in Fig.~\ref{fig:histogram-array} that the ensemble of
Eq.~(\ref{eq:low-freq-gibbs}) convincingly reproduces dynamical results after
1000 cycles.
Fruthermore, the mean energy density derived from Eq.~(\ref{eq:betarf-stat})
agrees well with the dynamics for $\tau\gtrsim 5$, see
Fig.~\ref{fig:beta-comp}.
These results underpin our main insight: at low and intermediate frequencies,
although being applied only locally, the driving gradually affects the entire
system as the reservoir synchronises with the system proper and an equilibrium
state with low net energy absorption emerges \cite{long-paper}.
This collective behaviour, which is mediated only by short-range interactions,
is in stark contrast with the conventional notion of a thermal reservoir as a
reversible heat sink, which is realised here only in the high-frequency regime.

\addcontentsline{toc}{section}{Perspectives}
\paragraph{Perspectives.--}
Our results provide strong evidence that effective equilibrium states may be
stroboscopically established in a periodically driven open many-body system
over a wide range of frequencies.
The limits $\omega \to 0$ and $\omega \to \infty$ match physical intuition, and
we have demonstrated that even leading-order correction in $\omega^{-1}$ and
$\omega$ respectively are sufficient to account for all but a small frequency
window around $\tau \approx 2$.
This crossover regime may still be described by a non-equilibrium
steady state featuring persistent currents.
A natural mechanism for generating such currents would be an emerging phase lag
in the synchronisation that governs the low-frequency regime.
Hence, an increasing inability of the system to respond to the drive could be
the physical origin for the breakdown of the low-frequency ensemble which, on
dimensional grounds, must occur at $\omega/J \sim 1$.

The existence of a rotating frame is contingent on our choice of the drive.
We nonetheless generically expect relaxation to an instantaneous Gibbs ensemble
in the quasi-static limit $\omega \to 0$.
The instantaneous Hamiltonian then plays the role of an adiabatic invariant,
and it would be interesting to pursue a perturbative construction of analogous
invariants away from this limit.
Although numerically accessible, the analytic construction of such ensembles is
likely not straightforward.
However, by numerically evaluating the response functions entering
fluctuation-dissipation relations, it is in principle possible to verify that a
system is in equilibrium without specific knowledge of its state.

There remain many interesting questions about the relationship between the
convergence of the high-frequency expansions and heating and the role of
effective conservation laws.
In addition, it would be instructive to explore how the physics that we have
uncovered here survives or is modified in higher dimensions, where the same
theoretical manipulations are in principle possible.
\addcontentsline{toc}{section}{Data access statement}
\paragraph*{Data access statement.--}
The source code used for all simulations, and all data used in figures, is
freely available at \url{https://github.com/tveness/spinchain-papers}.
\addcontentsline{toc}{section}{Acknowledgements}
\begin{acknowledgements}
\paragraph*{Acknowledgements.--}
We thank Anatoli Polkovnikov for suggesting the research problem and for
thoughtful discussions.
TV is grateful for hospitality at Boston University during the preparation of
this manuscript.
KB acknowledges support from the University of Nottingham through a Nottingham
Research Fellowship. This work was supported by the Medical Research Council
  [grant number MR/S034714/1]; and the Engineering and Physical Sciences Research
Council [grant number EP/V031201/1].
\end{acknowledgements}
\bibliography{ttt}

\end{document}